\documentclass[twoside]{procinasan}

\input{procinasan.def}
\usepackage{textcomp}

\begin{document}

\def\teff{$T_{\rm eff}$}
\def\lgg{log\,${g}$}

\pagebreak

\thispagestyle{titlehead}

\setcounter{footnote}{0}
\setcounter{equation}{0}
\setcounter{section}{0}
\setcounter{figure}{0}
\setcounter{table}{0}

\markboth{Romanovskaya et al.}{Fundamental parameters of Ap-star HD~108662}

\titlen{FUNDAMENTAL PARAMETERS OF Ap-STAR HD~108662}
{Romanovskaya A.M.$^{1}$, Ryabchikova T.A.$^1$, Shulyak D.V.$^2$}
{$^1$Institute of Astronomy of the RAS, Moscow, Russia\\
$^2$Max-Planck-Institute f\'ur Sonnensystemforshung, G\'ottingen, Germany}

\abstre{We present the results of a self-consistent spectroscopic analysis of the atmosphere of Ap-star HD~108662 
based on high resolution spectrum and low resolution spectrophotometric observations. 
Magnetic field effects, such as Zeeman broadening and polarized line formation, were taking into account in the analysis
of spectral line profiles. 
We derived abundances of 24 chemical elements as well as the stratification of Fe~--~an element which is the main contributor to the line absorption
in the visible wavelengths. Another abundant chemical element~--~Chromium~--~was found to be distributed homogeneously in the atmosphere of the star. 
From our analysis we determined the following fundamental parameters of HD~108662: \teff=10212~K, \lgg=4.0, $R/R_{\odot}$=2.09 and log$(L/L_{\odot})$=1.63. 
The estimated rotational velocity of the star is $v\,sini = 20.4$ km/s and the strength of the surface magnetic field is $B_s = 3300$~G.}

\selectlanguage{english}

\addcontentsline{toce}{subsection}{{\em Romanovskaya A.M., Ryabchikova T.A., Shulyak D.V.\/} The fundamental parameters of Ap-star  HD~108662}

\selectlanguage{english}

\baselineskip 12pt

\section*{Introduction}

The determination of fundamental parameters of stars  (\teff, \lgg, $R/R_{\odot}$, and log$(L/L_{\odot})$)  is one of the main goal of astrophysics. 
For magnetic chemically peculiar (Ap) stars the determination of their parameters is complicated due to the anomalous abundances of their atmospheres, 
which requires special methods for analysis tools. Exact values of the fundamental parameters are necessary 
for determining the evolutionary states of stars and their positions on the Hertzsprung-Russell diagram. 
Furthermore, accurate fundamental parameters are required for testing theoretical models of pulsations in magnetic stars, 
for which the knowledge of, e.g., accurate radii is essential \cite{2013MNRAS.436.1639C}.

A mechanism explaining the development of chemical anomalies was proposed by Michaud \cite{1970ApJ...160..641M}, 
according to which the separation of chemical species in the star's atmosphere results from the diffusion of atoms and ions of a chemical element 
under the combined action of several forces. The main ones are the stellar gravity which is directed toward the center of the star, 
and radiation pressure which pushing 
particles into the outer atmosphere layers. Since the diffusion separation rates are very small, the diffusion mechanism requires the stability 
of the star's atmosphere, i.e. primarily the lack of any turbulence. The presence of global magnetic fields in Ap stars stabilizes the atmosphere
and hence creates conditions for the diffusion of chemical elements and drives vertical gradients of abundances (stratification).

\section*{The determination of atmospheric parameters}
The star HD~108662 (17 Com) is the magnetic chemical peculiar star of A0p spectral class. 
The fundamental atmospheric parameters were determined by comparing the observed energy distribution against the theoretical one. 
Since the outgoing stellar radiation depends on the atmospheric absorption, 
it is necessary to correctly account for the peculiar atmospheric abundances as well as
the stratification of elements, especially those that make the largest contribution to the opacity. 
In this study, we utilize an iterative method to determine the fundamental parameters and abundances (including stratification).
It consists of subsequent refinement of model structure and spectroscopically derived abundances until the agreement between
predicted and observed spectra is achieved.
This approach was successfully used to study the atmospheres of many Ap stars \cite{2009A&A...499..851K, 2013A&A...551A..14S, 2019MNRAS.488.2343R}.

We used the observations from the ESPaDOnS archive for the spectroscopic analysis of the star\footnote{http://www.cadc-ccda.hia-iha.nrc-cnrc.gc.ca/en/cfht/}. 
The resolving power was  R=$\lambda/\Delta\lambda$ = 65000 in the spectral range 3700--10000~\AA. 
Near UV and optical spectrophotometric data were taken from the Adelman catalog \cite{1989A&AS...81..221A} and in the near IR range from the 2MASS catalog 
(2Micron All-Sky Survey -- \cite{2003yCat.2246....0C}).

As a first approximation, abundances and stratification were determined using atmospheric parameters taken from the literature. 
The initial calculation of the atmospheric model was carried out using ATLAS9 model atmosphere code \cite{1993KurCD..13.....K} 
according to the results by \cite{1996AstL...22..815S} (\teff=10300~K, \lgg=4.3). Then the model was recalculated with new refined parameters 
using the \textsc{LLmodels} code \cite{2004A&A...428..993S} which can account for the individualized abundances and vertical stratification
of chemical elements. The synthetic spectrum was calculated taking into account the magnetic field effects by using the Synmast code \cite{2007pms..conf..109K}, 
and the line transition parameters for calculating the synthetic spectrum were taken from the Vienna database VALD3 \cite{2015PhyS...90e4005R}. 
In addition, for a number of least blended lines, the equivalent width analysis method was used to infer abundances of chemical elements
with the help of the WidSyn code \cite{2013A&A...551A..14S}
which also takes into account the magnetic field effects in the formation of spectral lines.

The magnetic field was estimated from the differential magnetic broadening of the magnetically sensitive lines 
of Fe~II at 6149~\AA\ and 8352~\AA\ with Lande factors 1.35 and 1.28, respectively. 
For any configuration of the magnetic field, these two lines are split into two components. 
Therefore, they are most often used to estimate the magnitude of the magnetic field modulus $B_s$ averaged over the surface of the star. 
The projected rotational velocity was estimated simultaneously with the magnetic field: $v\,sini = 20.4(4)$ km/s. 
We measured the magnetic field value of $B_s = 3300(150)$~G (the numbers in parentheses are the errors in the last digits).

Starting for an initial model atmosphere, we derived abundance and stratification and computed a new grid of models. We then compared predictions of these
models against low-resolution spectrophotometric observations calibrated to absolute flux units, thereby refining the atmospheric \teff\, and \lgg.
With the known parallax, we also derived the stellar radius. Then, we used these atmospheric parameters to compute new synthetic spectrum and carried
out spectroscopic analyzes again. The whole process was repeated until the stellar parameters converged to some stable values.
For HD~108622 we needed four iterations to obtain final best-fit parameters.

\begin{figure}[t]
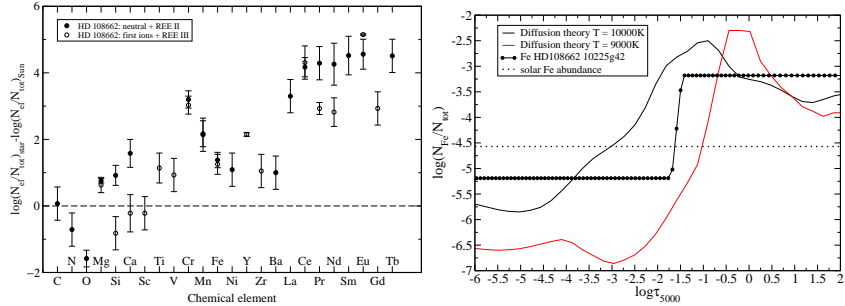

	\centering
	\includegraphics[width=0.49\textwidth, clip]{HD108662_abundance.eps}
	\includegraphics[width=0.49\textwidth, clip]{Fe_HD108662_str.eps}
	\caption{Element abundances (left panel) and stratification of Fe (right panel) in the atmosphere of the HD~108662.
The average values of element abundances over all lines are presented as $log(N_{el}/N_{tot})$. The red and black lines show the theoretical 
Fe stratification for models 9000g40 and 10000g40 \cite{2005JRASC..99T.139L, 2009A&A...495..937L}. The iron abundance in the solar atmosphere 
is shown by a dotted line.}
	\label{stratif}
\end{figure}

Figure~\ref{stratif} (left panel) shows mean element abundances in the atmosphere of HD~108662 relative to the solar photospheric values. 
The solar abundances were taken from \cite{2015A&A...573A..25S, 2015A&A...573A..26S, 2015A&A...573A..27G}. 
The derived abundances are typical for Ap stars and are characterized mainly by excesses up to an order of magnitude of abundances of iron peak elements 
and very significant excesses of rare-earth elements in comparison with solar values. There is also a significant He deficiency in the star. 
Helium lines are practically not observed, therefore we used the limiting value of -4.05~dex (predicted by the diffusion theory)
in our model atmosphere calculations. We determined the abundances of 24 elements, and for 10 of them we were able to analyze 
lines originating from two stages of ionization.

Stratification of chemical elements depends on the \teff, \lgg, and $B_s$ of the star \cite{2009A&A...495..937L}. 
The stratification profile can be described by a step function with four parameters: element abundances in the upper and lower atmospheric layers, 
the position of the abundance jump and the width of this jump \cite{2005A&A...438..973R, 2009A&A...499..879S}. 
To analyze the Fe stratification we used the DDaFit code \cite{2005A&A...438..973R}. 
For this, 15 Fe~I/II lines with different equivalent widths and excitation energies $E_i$ were selected to probe the formation of lines at different 
atmospheric depths. The derived Fe stratification in the atmosphere of HD~108662 is shown in Fig.~\ref{stratif} (right panel). 
Theoretical calculations of the Fe diffusion \cite{2005JRASC..99T.139L} show that, as the temperature of the stellar atmosphere increases,
the position of the abundance jump shifts to the upper atmosphere layers and the difference in the abundance between lower and upper 
layers decreases. Thus, our results agree very well with the theoretical predictions. 
It was not possible to derive the stratification of another very abundant element of the iron group, Chromium, 
because our spectra do not contain suitable unblended lines in the second stages of ionization in a wide range of excitation energies.

As was mentioned above, it took us four iterations for HD~108662 to obtain final fundamental parameters, 
with recalculation of abundances and stratification at each iteration. We fit the theoretical flux for the corresponding atmospheric model 
to the observed energy distribution using the parallax $\pi$ = 13.5382(2245)~mas which we took from the GAIA DR2 catalog \cite{2018A&A...616A...1G}.
Since HD~108662 is a bright star at a close distance from the Sun, interstellar reddening was not taken into account, which is consistent with 
the limiting value E(B--V) = 0.002(14) obtained from the interstellar absorption maps in \cite{2014A&A...561A..91L}.    

\begin{figure}[t]
	\centering
	\includegraphics[width=0.85\textwidth, clip]{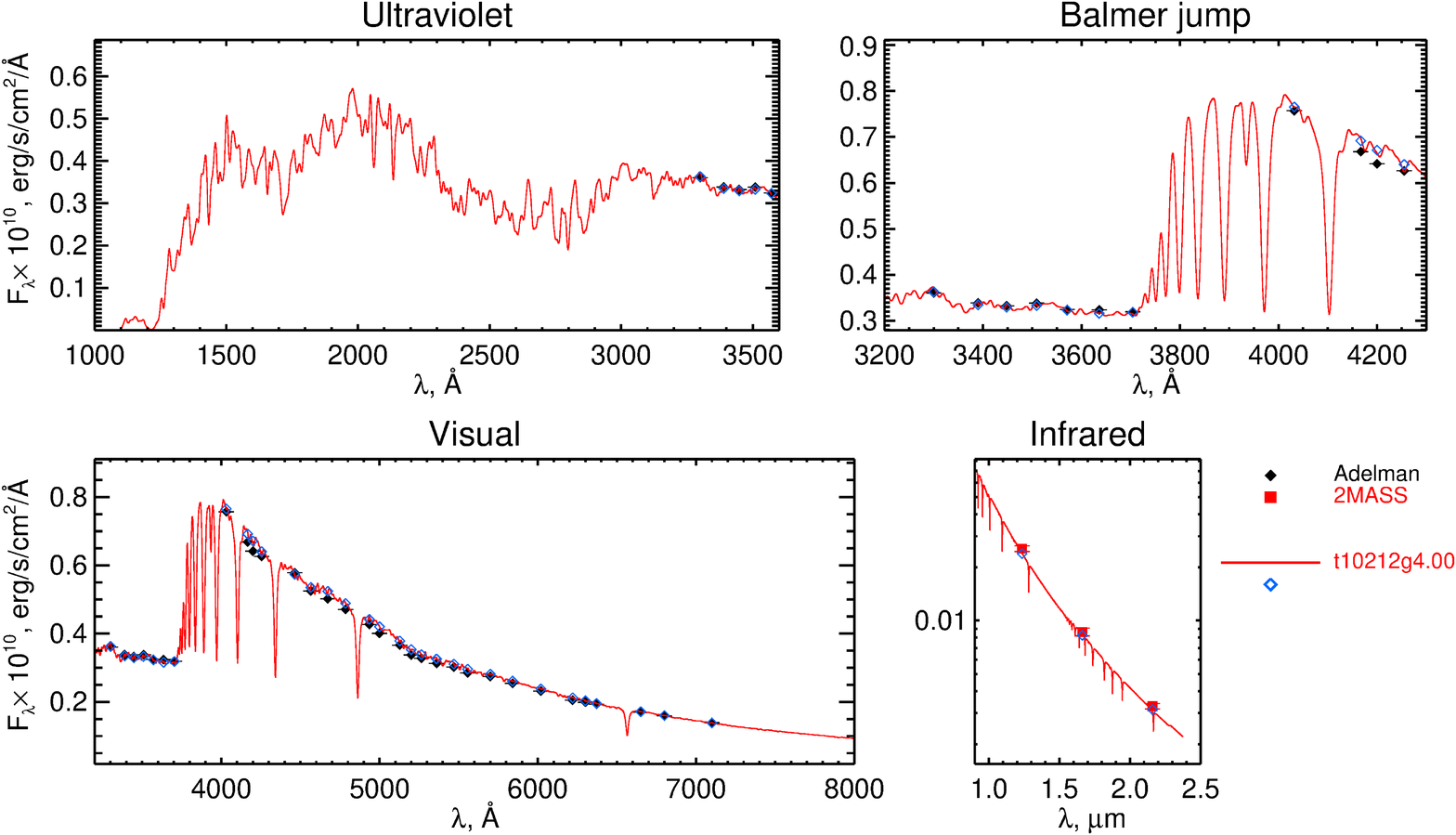}
	\caption{Comparison between the observed energy distribution (filled squares and rhombs) with the theoretical one calculated using 
    the \textsc{LLmodels} code (red solid line) for the atmosphere model of HD~108662 with parameters \teff=10212~K, \lgg=4. 
    Open diamonds show the theoretical flux after convolution with the corresponding filters used in observations.}
	\label{sed}
\end{figure}

\begin{figure}[hbt!]
	\centering
	\includegraphics[width=0.80\textwidth, clip]{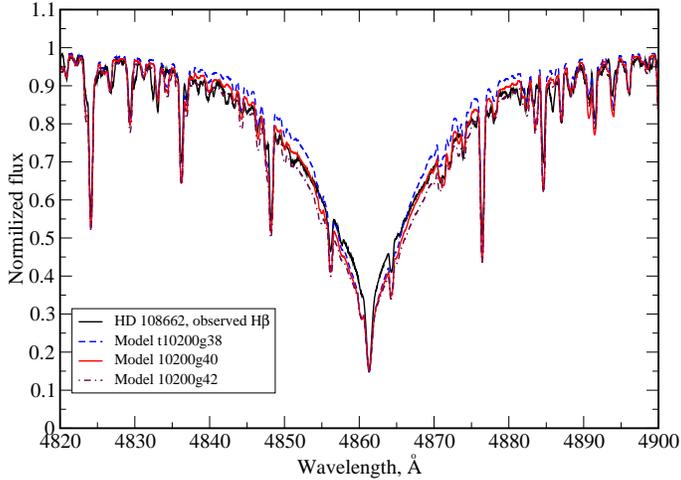}
	\caption{Comparison between the observed and predicted H$\beta$ line profile for models with \teff\ = 10200~K and with \lgg\ = 3.8 (blue dashed line), 
    4.0 (red solid line), 4.2 (brown dashed line).}
	\label{hbeta}
\end{figure}

The best agreement between model and observations was obtained for a model with parameters \teff\ = 10200(100) K, \lgg\ = 4.00 and a star radius $R/R_{\odot}$ = 2.09(10). 
The radius error was calculated taking into account the error on the parallax. The value of \lgg\ could not be robustly determined from the fit to the observed
energy distribution and therefore was chosen to be \lgg=4.00 as the most suitable for fitting the hydrogen lines profiles (Fig.~\ref{hbeta}). 
A comparison of the theoretical energy distribution with the observed one is shown in Fig.~\ref{sed}. 
The obtained effective temperature and radius of the star allow us to estimate the star's luminosity log$(L/L_{\odot})$ = 1.63(06).

\begin{figure}[hbt!]
	\centering
	\includegraphics[width=0.80\textwidth, clip]{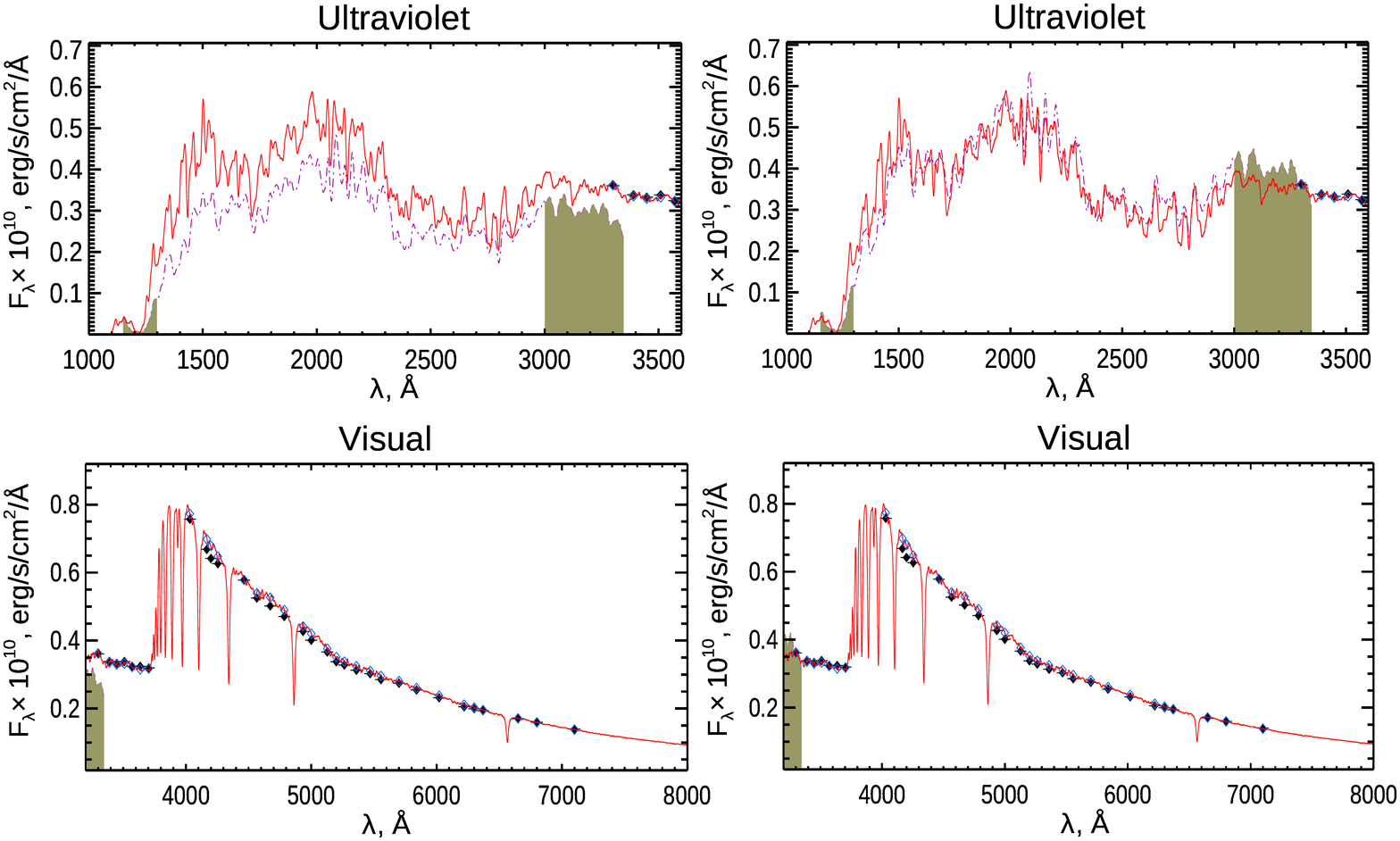}
	\caption{Comparison between the observed and predicted energy distributions using the \textsc{LLmodels} (red solid line) for the atmospheric model with parameters \teff=10212~K, \lgg=40. 
    The line colors are the same as in Fig.~\ref{sed}. IUE observations are indicated by a dash-dotted violet line. 
    The spectral range between UV and optical observations is marked in green. The original IUE data is shown on the left, UV fluxes on the right, 
    taking into account the scaling factor of 1.32 (see text).}
	\label{sed_comp}
\end{figure}

It should be noted that for HD~108662, there are observations in the ultraviolet region 1900--3000~\AA\ obtained with the International Ultraviolet Explorer telescope (IUE)\footnote{http://archive.stsci.edu/iue/}. 
However, when constructing the spectral energy distribution in a wide wavelength range, it turned out that there is a noticeable disagreement 
in the narrow spectral range between UV and optical observations (see Fig.~\ref{sed_comp}, left panels). A good agreement between 
the observed and theoretical energy distribution could only be obtained assuming a scaling factor of 1.32 for the IUE flux. 
Note that our predicted flux fits the observed one in all other wavelengths (visual and IR) as well as the observed hydrogen line profiles 
(Fig.~\ref{sed_comp}, right panels). It is thus possible that inaccuracies in the calibration of UV fluxes might have taken place due to, e.g.,
the relatively high brightness of HD~108662 (V$^m$=5.236) and/or the star being positioned incorrectly during some of the IUE observations which 
resulted in a noticeable flux loss through the instrument slit.

\section*{Conclusion}
In this work we carried out a detailed analysis of the atmosphere of the chemically peculiar star HD~108662
using high resolution spectroscopic and low resolution spectroscopic observations calibrated to absolute units.
The values of the rotation velocity and the strength of the magnetic field 
were found to be 20.4~km/s and 3300~G, respectively. We also derived the mean values of element abundances in the atmosphere of the star,
as well as the inhomogeneous vertical distribution of iron. The latter shows a very good agreement with the predictions of the diffusion models.

By taking into account the anomalies in the abundances and elemental stratification by successive iterations in our spectroscopic analysis, 
we determined the fundamental atmospheric parameters of the star to be \teff=10212~K, \lgg=4, R=2.09R$_{\odot}$, log$(L/L_{\odot})$=1.63.

In this work, we made use of the VizieR and VALD3 databases. Our study was funded by RFBR, project number \textnumero19-32-90147.

\small
\bibliography{Romanovskaya_ref}
\normalsize
\end{document}